Metal-insulator transition and giant anisotropic magnetoresistance in ultra thin (Ga,Mn)As.

R.R. Gareev[1], A. Petukhov[2], M. Schlapps[1], M. Doeppe[1], J. Sadowski[3], M. Sperl[1], and W. Wegscheider[1]

*[1] - Institute of Experimental and Applied Physics, University of Regensburg, Universitaetstrasse 31, 93040 Regensburg, Germany; [2] - South Dakota school of Mines and Technology, 501 East Saint Joseph Street, Rapid City, SD 57701, USA; [3] - MAX-Lab, Lund University, 22100 Lund, Sweden*

MBE-grown, 5 nm-thick annealed $Ga_{0.95}Mn_{0.05}As$ films with $T_c \sim 90K$ demonstrate transition from metallic to insulating state below $T_o \sim 10K$, where sheet resistances $R_{sh} \sim h/e^2$ and both longitudinal $R_{xx}$ and transverse $R_{xy}$ components become comparable. Below metal-insulator transition we found giant anisotropic magnetoresistance (GAMR), which depends on orientation of magnetization to crystallographic axes and manifests itself in positive magnetoresistance near 50% for $R_{xx}$ at T=1.7K, H//[110] crystallographic direction and parallel to current in contrast to smaller and negative magnetoresistance for H//$[1\bar{1}0]$ direction. We connect GAMR with anisotropic spin-orbit interaction resulting in formation of high- and low-resistance states with different localization along non-equivalent easy axes.

Nano-scaled (Ga,Mn)As magnetic semiconductor films are expected to possess quite unusual magnetic and transport properties. Relatively weak exchange interactions can lead to collapse of global ferromagnetic (FM) ordering with establishment of localization effects for films thinner than several nanometers.[1] Theoretical considerations indicate that the local FM ordering can be stabilized by double exchange or by formation of magnetic polarons.[2,3] Depletion of carriers close to interfaces can additionally favour localization.[4,5] Recent experiments on ultra thin (Ga,Mn)As films have confirmed the insulating-type transport *via* localized holes as well as a weakened FM exchange.[6]

Studies of the anisotropic magnetoresistance (AMR) are of special interest due to anisotropic spin-orbit interaction (SOI) in (Ga,Mn)As. For metallic (Ga,Mn)As films AMR is small and positive.[7] However, by approaching the metal-insulator transition (MIT) from metallic side nano-scaled films exhibit an increased AMR, which was related to a weak localization (WL).[8] As demonstrated earlier, laterally constricted (Ga,Mn)As films and nanowires, as well as sandwich (Ga,Mn)As/GaAs/(Ga,Mn)As structures exhibit an enhanced magnetoresistance.[9-13] Tunnelling anisotropic magnetoresistance (TAMR) effect, which has been observed in sandwich structures manifests itself in dramatic changes of current-perpendicular-to plane resistance upon switching of the magnetization *M* between non-equivalent easy axes. The TAMR was connected with anisotropic SOI in strained (Ga,Mn)As.[11] Deep in the insulating state the coulomb blockade AMR arises for comparable anisotropy in the chemical potentials and single electron charging energies[13].

The behaviour of resistance in ultra thin films for sheet resistances $R_{sh}$ comparable to the resistance quantum $R_0 = h/e^2 = 25.8$ k$\Omega$ we connect with the MIT. Close to the MIT from insulating side AMR enhances and becomes dependent on the orientation of *M* with respect to crystallographic axes. In contrast to TAMR the giant anisotropic magnetoresistance (GAMR) in our ultra thin films we observed in the lateral geometry of current.

We prepared ultra thin (Ga,Mn)As films with the nominal content of Mn near 5% at the growth temperature $T_g$ ~250 °C using electron–beam evaporation. In order to reduce the out-diffusion of Mn we employed a 5nm-thick $Al_{0.7}Ga_{0.3}As$ barrier layer grown ontop GaAs/GaAs(001) buffer. In order to increase effective concentration of holes $p$ we used post-growth annealing at T~$T_g$. The optimal annealing time τ corresponded to the lowest values or resistance (highest $p$). For instance, for the 5nm-thick films we found τ ~0.5 hour. Magnetic properties were studied using Quantum Design superconducting quantum interference device (SQUID). The Curie temperature $T_c$ was registered by the abrupt drop of *M*. The prepared films were patterned into Hall bars by using optical lithography. The width of current paths as well as the separation between contacts was 50 μm and, thus, longitudinal resistance $R_{xx}$ corresponded to the sheet resistance $R_{sh}$.

Applying alternating current we measured both longitudinal $R_{xx}$ and transverse $R_{xy}$ components of resistance. Data were taken by four-probe method using standard lock-in amplifier with the current path oriented along the [110] direction. The amplitude of the applied current *I* was ~100nA.

In Fig.1a) we present typical *M*(T) dependencies, which illustrate the enhancement of both *M* and $T_c$ by annealing. Dependences of $T_c$ on thickness *t* are presented in the inset in Fig. 1a). The results for films with and without (Al,Ga)As diffusion barrier i) and before and after annealing ii) are presented. It is seen that without the barrier only 5 nm-thick films display FM ordering. By using such a barrier we realized FM ordering for thinner 3 nm films. The combined effect of barrier and annealing leads to magnetic ordering even for *t*=2 nm. The Curie temperature $T_c$ shows a continuous increase with *t* and reaches $T_c$~ 90K for 5 nm-thick annealed films with diffusion barrier.

Below we describe transport in 5 nm-thick films, which exhibit $R_{xx}=R_{sh}\sim R_0$ and a well-defined MIT below $T_0\sim 10K$. In the metallic state we found p-type conductivity with $p\sim 4*10^{20}cm^{-3}$ using procedure described elsewhere.[14] Fig. 1b) shows typical for metals $R_{xx}(H)$ and $R_{xy}(H)$ dependences. The magnetization-dependent Hall component $\Delta R_{xy}\sim 260\ \Omega$ remained stable above $T_0$ with pronounced drop close to $T_c$. The corresponding positive magnetoresistance did not exceed 1%.

The zero-field temperature dependences of $R_{xx}$ and $R_{xy}$ are presented in Fig. 2. In accordance to previous experiments with ultra thin (Ga,Mn)As[15], maximum of $R_{xx}$ is located at higher temperatures ($T^{**}\sim 120K$) compared to $T_c\sim 90K$. Higher values of $T^{**}$ are probably connected with residual local magnetic ordering, which can lead to an increased scattering above $T_c$.[16] Between $T^{**}$ and $T^*\sim 50K$ we found metallic behaviour, which is followed by a gradual increase of resistance between $T^*$ and $T_0$. We connect this increase with WL. Actually, recent studies of phase coherence in (Ga,Mn)As give $L_\Phi\sim 100nm$ at $T=100mK$ and power-law temperature dependence of $L_\Phi\sim T^{-1/2}$.[14] From these data we extrapolate that $L_\Phi\sim t$ at $T\sim 10K$ and, thus, phase-coherent hole transport becomes two-dimensional (2D). Finally, near $T_0$, where $R_{sh}$ becomes comparable to the resistance quantum $R_0=h/e^2$ we found strong increase of resistance, which we ascribe to MIT. These values are in a good accordance with the Thouless criterion for minimal metallic conductivity $\sigma_{min}\sim 1/R_0=e^2/h$ for 2D non-interacting systems.[17] The thermal diffusion length $L_T\sim 200nm$ at $T\sim 20mK$ also demonstrates $T^{-1/2}$ dependence.[18] Accordingly, $L_T\sim t$ for $T\sim 10K$ and diffusive transport is 2D even for interacting carriers. Thus, for both localization mechanisms (WL and e-e interactions) transport is 2D at $T\sim 10K$ and MIT occurs at $R_{sh}\sim R_0$. We found that in the insulating state both $R_{xx}$ and $R_{xy}$ components become comparable (see Fig.2). Increased values of $R_{xy}$ component in the insulating state can be connected with the mesoscopic character of transport between localized regions.

Temperature dependence of conductivity close to MIT can be described by the power-law $\sigma_{xx}=\sigma_0+bT^n$, where $\sigma_0$ is the conductivity at zero temperature and n=1/2 for 2D metals with WL. The inset in Fig.2 demonstrates $\sigma(T^{-1/2})$ dependence close to the MIT, which is linear above $T_0$ and in accordance with 2D transport. At $T_0$ we found a well-defined kink in conductivity. Below $T_0$ (region B) temperature coefficient b increases and $\sigma_0$ changes its sign to negative, which is characteristic for insulators. The observed MIT is not accompanied with transition from a power-law to an exponential $\sigma(T)$ dependence as expected for disorder-induced Anderson-Mott MIT.[19] We believe that for our films near MIT the Fermi level is situated close to the mobility edge and band edge conduction[20] dominates. In the region C with decreasing temperature the longitudinal resistance demonstrates only a slight increase. This "frozen" state we relate to formation of localization gap due to unscreened e-e interactions.

In Fig.3a we present magnetoresistance for the longitudinal component $R_{xx}$ at T=1.7K for three different orientations of *H* to crystallographic axes. The most dramatic changes of $R_{xx}$ we observed for *H//I//[110]*. For this orientation we found GAMR effect, which we define as the relative difference between stable high-resistance $R_{HR}$ and low-resistance $R_{LR}$ states (GAMR = $(R_{HR}-R_{LR})/R_{HR}*100\%$). The GAMR is positive and near 50% at T=1.7K. The established GAMR effect is not connected with standard AMR for isotropic conductors, where $R_\perp(I\perp H)>R_{//}(I\perp H)$, effect is small and does not depend on the angle between *M* and crystallographic directions.[21] In contrast, the GAMR shows opposite sign ($R_\perp<R_{//}$) and depends on orientation of *M* to crystallographic axes. Alternatively, we connect the GAMR with the crystalline component of AMR resolved in ultra thin metallic (Ga,Mn)As close to the MIT.[22] We emphasize that GAMR, which we established below MIT is at least one order of magnitude stronger compared to AMR. We connect magnetization behaviour with non-equivalency of [100] and [010] biaxial easy axes in compressively-strained (Ga,Mn)As ultra thin films caused by uniaxial anisotropy. The uniaxial easy axis is aligned along [100] direction (Fig.3a). Then positive GAMR can be explained by switching and rotation of *M* between two stable easy axes and formation of low resistance and high resistance states along uniaxial easy and hard axes, accordingly. We note that GAMR is observable close to the saturation field. Different resistance states can be connected with the anisotropic extent of bound hole states in presence of strains and spin-orbit interaction.[23] Then the MIT can proceed as percolation of extended two-dimensional bound hole states.

Concluding, we prepared ultra thin ferromagnetic (Ga,Mn)As films with well-defined MIT. Close to the MIT 5nm-thick films exhibit sheet resistances near resistance quantum and comparable longitudinal and transverse components of resistance related to two-dimensional transport. Below the MIT we established the GAMR effect for the planar geometry of current, which is reflected in stable states with different resistance. The positive GAMR reaches 50% for the $R_{xx}$ component at T=1.7K for magnetic field parallel to current and [110] crystallographic direction. We connect GAMR with anisotropic spin-orbit interaction in presence of the uniaxial anisotropy. The abrupt and strong changes of resistance between stable high- and low-resistance states can be useful for planar anisotropic magnetic switches.

Authors thank Prof. Dieter Weiss for stimulating discussions and appreciate the financial support from the Project SFB 689.

Figures and figure captions:

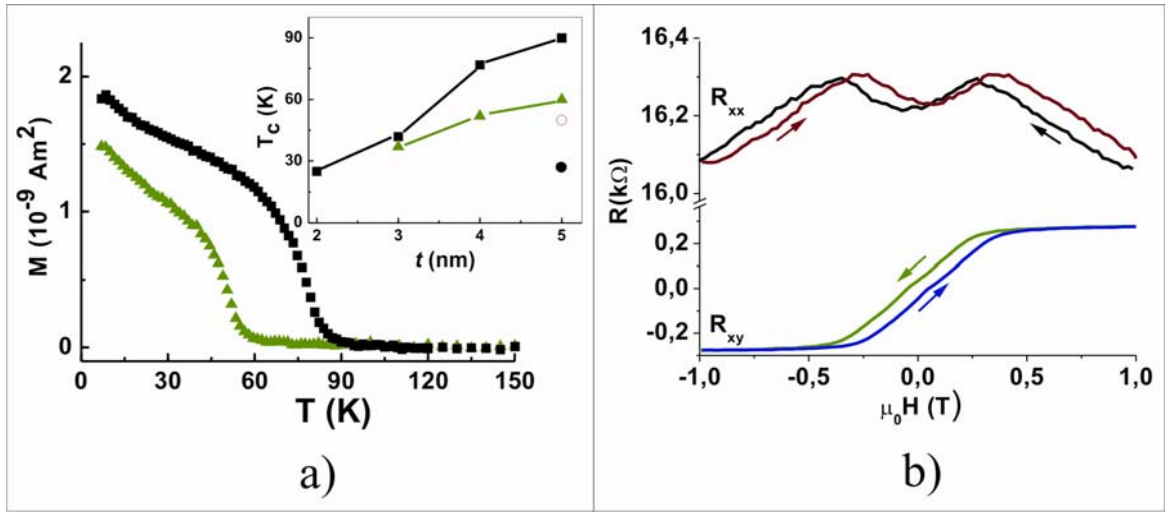

**Fig.1.** Optimization of magnetic and transport properties of ultra thin $Ga_{0.95}Mn_{0.05}As$ films. In Fig. 1a) is shown the temperature dependence of magnetization $M$ upon cooling down in magnetic field $H$=100 Oe applied along the [110] direction before annealing (triangles) and after annealing (quadrats) of the 5 nm-thick film grown on $Al_{0.7}Ga_{0.3}As$ buffer. The inset demonstrates the dependence of the Curie temperature $T_c$ on the film thickness $t$ under different preparation conditions without barrier: before (filled circle) and after annealing (open circle) i); grown on top of barrier: not annealed (triangles) and after annealing (quadrats) ii). Fig.1 b) shows magnetoresistance dependences for $R_{xx}$ and $R_{xy}$ components in the metallic state taken at T=50K in the Hall geometry of $H$ for 5 nm-thick films with optimized properties. Arrows indicate directions of the field sweep.

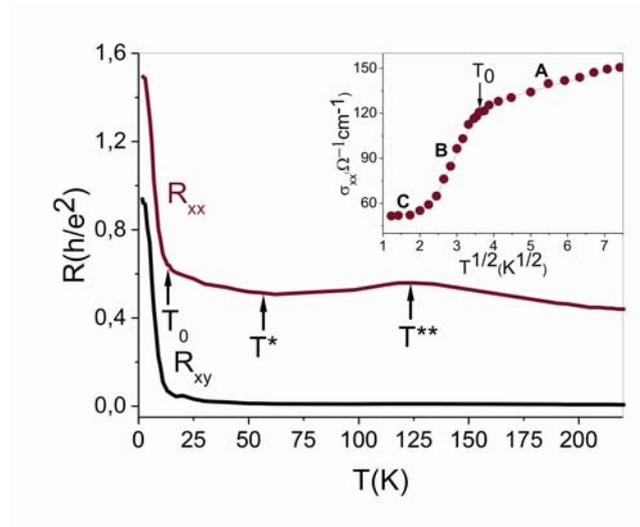

**Fig.2**. Dependence of resistance on temperature for $R_{xx}$ and $R_{xy}$ components upon cooling down without magnetic field. Arrows indicate positions of maximum of resistance $T^{**}$ near $T_c$, minimum of resistance in the metallic state ($T^*$) and temperature $T_0$ of the MIT. The inset demonstrates longitudinal conductivity $\sigma_{xx}$ *versus* $T^{1/2}$ dependence, where $\sigma_{xx} = 1/(R_{xx}*t)$. Arrow indicates temperature of the MIT. Symbols A, B and C in the inset mark regions with different transport behaviour. The dependence $\sigma_{xx}(T^{1/2})$ follows the linear law $\sigma_{xx} = \sigma_0 + bT^{1/2}$, where $\sigma_0(A) \sim 100\,\Omega^{-1}\mathrm{cm}^{-1}$; $b \sim 7\,\Omega^{-1}\mathrm{cm}^{-1}\,\mathrm{K}^{-1/2}$ and $\sigma_0(A) \sim -10\,\Omega^{-1}\mathrm{cm}^{-1}$; $b \sim 53\,\Omega^{-1}\mathrm{cm}^{-1}\,\mathrm{K}^{-1/2}$ for regions A (metallic) and region B (insulating), correspondingly.

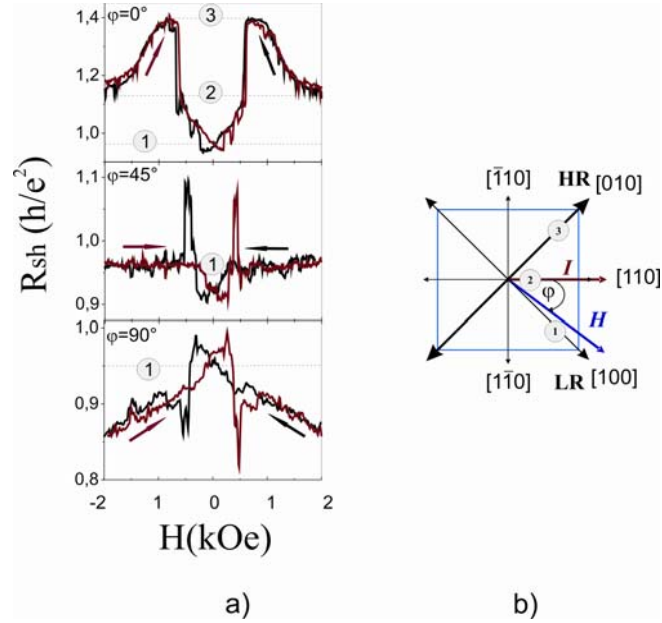

**Fig.3**. Giant anisotropic magnetoresistance in $Ga_{0.95}Mn_{0.05}As$ with thickness 5nm at temperature T=1.7K. Longitudinal $R_{xx}$ component of resistance *versus H* for different angles φ: 0°, 45° and 90° between magnetic field *H* and current *I*//[110] direction (Fig.3a). Arrows indicate direction of the field sweep. The orientations of magnetic field, uniaxial hard axis and biaxial easy axis are shown in Fig.3b) by thick blue, thick black and thin black arrows, respectively. Behaviour of resistance reflects orientation of *M* vector to crystallographic directions. In the remanent state magnetization is aligned along biaxial easy axis 1. For *H* along biaxial hard axis (φ =0°) vector of *M* passes through two stable states with different resistance: low-resistance (LR) state 1- along biaxial easy axis and high-resistance (HR) state 3 along uniaxial hard axis. Starting from the saturation field *H* (state 2) the *M* vector coherently rotates (2-3), switches by passing biaxial hard axis (3-2), and continuously rotates to LR biaxial easy axis (2-1).